\documentstyle{article}
\newcommand{\np}[2]{{\em Nucl.\ Phys.\ }{\bf #1}{(#2)}}
\newcommand{\pr}[2]{{\em Phys.\ Rev.\ }{\bf #1}{(#2)}}

\newcommand{\prsl}[2]{{\em Proc.\ Roy.\  Soc.\  A\  (London)\ }{\bf #1}{(#2)}}
\newcommand{\pps}[2]{{\em Proc.\  Phys.\  Soc.\  }{\bf #1}{(#2)}}
\newcommand{\pl}[2]{{\em Phys.\ Lett.\ }{\bf #1}{(#2)}}
\newcommand{\jmp}[2]{{\em J.\ Math.\ Phys.\ }{\bf #1}{(#2)}}
\newcommand{\jmps}[2]{{\em J.\ Math.\& \ Phys.\  Sci. \  (India) \ }{\bf
#1}{(#2)}}

\newcommand{\mpl}[2]{{\em Mod.\ Phys.\ Lett.\ }{\bf #1}{(#2)}}
\newcommand{\ijmp}[2]{{\em Int.\ J.\ Mod.\ Phys.\ }{\bf #1}{(#2)}}
\newcommand{\rmp}[2]{{\em Rev.\ Mod.\ Phys.\ }{\bf #1}{(#2)}}
\newcommand{\apny}[2]{{\em Ann.\ Phys.\ (N.Y.)\ }{\bf #1}{(#2)}}
\newcommand{\jpa}[2]{{\em J. \ Phys. A\ }{\bf #1}{(#2)}}
\newcommand{\jpg}[2]{{\em J. \ Phys. G\ }{\bf #1}{(#2)}}
\newcommand{\ajp}[2]{{\em Am. \ J. \ Phys.\ }{\bf #1}{(#2)}}
\newcommand{\fp}[1]{{\em Forts.\ Phys.\ }{\bf #1}}
\newcommand{\nc}[2]{{\em Nuovo\ Cim. \ }{\bf #1}{(#2)}}
\newcommand{\ncs}[2]{{\em Nuovo\ Cim.\ Suppl. \ }{\bf #1}{(#2)}}
\newcommand{\yf}[2]{{\em Yadern.\ Fiz. \ }{\bf #1}{(#2)}}
\newcommand{\sjnp}[2]{{\em Sov.\ J.\ Nucl.\  Phys. \ }{\bf #1}{(#2)}}
\newcommand{\apa}[2]{{\em Acta\  Phys.\  Austr.\ }{\bf #1}{(#2)}}

\newcommand{\hj}[2]{{\em  Hadronic \ J. \ }{\bf #1}{(#2)}}
\def\theequation{\arabic{section}.\arabic{equation}}
\def\appendix{
\vskip 1cm
\par
\setcounter{equation}{0}
\def\theequation{A1.\arabic{equation}}}
\begin{document}
\begin{titlepage}
\large{\begin{center}
\hspace*{10cm} Preprint IFUNAM\\
\hspace*{10cm} FT-94-36\\
\hspace*{10cm} January 1994\\
\vspace*{25mm}
{\bf $2(2S+1)$- COMPONENT MODEL AND ITS CONNECTION \\ WITH OTHER
FIELD THEORIES  $^{\star}$}\\
\vspace{6mm} {\bf  Valeri  V. Dvoeglazov}$^{\,\dagger,\, \ddagger}$\\
\vspace*{3mm}
{\it  Departamento de F\'{\i}sica Te\'{o}rica, \,Instituto
de F\'{\i}sica,\\
Universidad Nacional Aut\'{o}noma de M\'{e}xico, \\
Apartado Postal 20-364, 01000 D.F. , MEXICO}\\
\end{center}
\vspace*{6mm}
\noindent
\begin{abstract}
\noindent
\large{

\noindent
This talk presents  the review of  forgotten but attractive formalism proposed
by Joos and Weinberg in the sixties for description of high-spin particles.
Problems raised in the recent works [Ahluwalia {\it et al.}] are discussed.
New results obtained  by the author in  his preceding  papers
["Hadronic J.", 1993, v. 16, No. 5, pp. 423-428;  No. 6, pp. 459-467;
Preprints IFUNAM FT-93-19, 24, 35] are reported.
In  {\it Appendix},  bibliography of publications related with mentioned
$2(2S+1)$-
component  formalism  is presented.}
\end{abstract}
\vspace*{35mm}
\noindent
KEYWORDS: high-spin particles, Lorentz group
representation, quantum electrodynamics,  Dirac oscillator\\
PACS: 03.50.De, 11.10.Ef, 11.10.Qr, 11.17+y, 11.30.Cp}\\
\noindent
-----------------------------------------------------------------------------------------\\

\vspace*{-5mm}
\footnotesize{
\noindent
$^{\star}$ Talk presented at the {\it  Escuela Latinoamericana de F\'{\i}sica
(ELAF'93). ''Fenomenolog\'{\i}a de las Interacciones Fundamentales''. Mar del
Plata, Argentina.
July 5-16, 1993} and at the {\it XVII Symposium on Nuclear Physics.
Oaxtepec, M\'exico. January 4-7, 1994}.\\
\noindent
$^{\dagger}$  Perm. address: {\it Dept. Theor.} \& {\it Nucl. Phys.,  Saratov
State
University  and Sci.} \& {\it Tech. Center for  Control and Use
of Physical Fields and Radiations,  Astrakhanskaya str. , 83, Saratov
RUSSIA}\\
\noindent
$^{\ddagger}$ Email: valeri@ifunam.ifisicacu.unam.mx,
dvoeglazov@main2.jinr.dubna.su}\\
\end{titlepage}

\newpage

\noindent




\section{Introduction}

Theoretical investigation of relativistic high-spin fields is very important
in the connection with the present experimental situation when
hadrons of spin up to 6  have been found~\cite{PDG}, see {\it Table I}. A
correct modelling of reactions which include high-spin states requires a
Lorentz covariant treatment
of these quantum fields.
\begin{center}

{\it Table \  I.} High-spin hadron resonances. \\

\vspace*{2mm}
\begin{tabular}{||c|c|c|c||}
\hline
\hline
Meson&$a_6(2450)$&$I^G=1^-$&$J^{PC}=6^{++}$\\
\hline
Meson&$f_6(2510)$&$I^G=0^+$&$J^{PC}=6^{++}$\\
\hline
Baryon&$\Delta(2420),H_{3,11}$&$I={3\over 2}$&$J^P=
{11 \over 2}^+$\\
\hline
Baryon&$N(2600),I_{1,11}$&$I={1 \over 2}$&$J^P=
{11 \over 2}^-$\\
\hline
\hline
\end{tabular}\\
\end{center}

\vspace*{1mm}

First  of all,  let me draw some preliminary statements connected with the
subject of my talk:

--- There is no exist a fully satisfactory formalism for high-spin fields.
Neither Rarita - Schwinger approach~\cite{rar},  in which wave function  is a
tensor of $S$th rank for bosons or a product of a Dirac spinor and a tensor of
$S$th  rank for fermions, nor Bargamann-Wigner approach~\cite{barg}, in which
wave function presents oneself a symmetric $2S$th order spinor,  provide
consistent field equations in the case of spin $S\geq1$ when gauge interaction
includes ($\partial_\mu\rightarrow \partial_\mu - ieA_\mu$).  The
Joos-Weinberg approach~\cite{Joos,Weinberg} considered in Section 2 is also not
free from difficulties.  E. g.,  in ref.~\cite{ahl}  it  was shown that
Weinberg equations have kinematically anomalous solutions and, moreover, suffer
from problem of interpretation negative energy solutions which have incorrect
parity. Analogously to other formalisms, the Joos-Weinberg equations do not
provide satisfactory mechanism for including  interaction,  as marked in ref.
[6a]. However, recent development of this formalism,  undertaken by
Ahluwalia {\it et al.}~\cite{ahl2}, namely, construction of a
Bargamann-Wightman-Wigner-type quantum field theory~\cite{Bargm}, in which a
boson and its antiboson appear with opposite relative intrinsic parity,
recasting Weinberg equations allowing for additional factor $\wp_{u,v}$ in
front of mass term,  gives more trust  in construction of adequate formalism
for description of high-spin hadron resonances.

---  In the book~\cite[p.77]{Hatf} it was written: 'The fields associated with
these particles,
$\phi$ and  $\psi$ (V. D.~: spinor and scalar), are not classically observable.
There is no
well-defined (measurable) classical limit or classical field theory to guide us
through
quantization. This is not true for the photon (free, massless, spin $1$
boson)... Yet, even though we have an experimentally well-defined classical
theory to guide us, the quantization of
electromagnetic field is by far the most delicate.'

 In fact,  such notions as a bispinor, a Lagrangian of spinor field, a
self-interaction potential etc. are required in relativistic quantum models
only.  Why should  we have the classical analog for electromagnetic field,
$A_\mu$, the 4- vector potential?

--- In the procedure of quantization of electromagnetic field it is necessary
to satisfy
simultaneously: a) requirement of positive energy density; b) Lorentz
condition; c) condition of transversality; d) relativistic covariance. As a
result, problem of indefinite metrics appears when quantization of
electromagnetic field. We need to assume that zero-component of vector
potential is anti-Hermitian. As mentioned in~\cite[p.93]{Corson}, 'the physical
meaning of so-called quantization on indefinite metric in the system Hilbert
space is obscure'.
There are still some attempts  of construction of electromagnetic field theory
without indefinite metrics~\cite{tirring}.

--- Well-known formalisms for description of massive vector particles (Kemmer
$\beta$- formalism, Proca  theory etc.) have no renormalizability. In order to
get the renormalizable theory with heavy vector bosons we need  implementing
the Higgs mechanism or  additional heavy vector particle~\cite{Hos} coupled to
the conserved current~\cite{sohrtoki}  or  using  non-local current
operator~\cite{Arno}.

--- It was observed by Tam and Happer~\cite{tam} that circularly polarized
laser beams
of opposite polarization repel each other, beams of equal polarization attract
in a medium
of sodium vapour.
This result can be interpreted in terms of  long-range spin-spin forces
resulting from exchange of massless axial vector gauge particles (in the first
order of perturbation theory), ref.~\cite{Naik}, without
applying to non-linear electrodynamics.

--- Introduction of axial magnetic current ($g_\mu=-g\bar \Psi\gamma_\mu
\gamma_5 \Psi$)
and axial-vector potentials permits one to avoid several difficulties in the
Dirac theory of magnetic monopoles, namely, it has not to resort to a
pseudoscalar magnetic charge; the quantum of  field's angular momentum is not
necessarily half integer~\cite{Nemes}.

--- Essential disagreement turns out between theoretical
prediction~\cite{pr11}:
\begin{equation}
\Gamma_{3}^{theor}(o-Ps)=7.038\,236(10)\, \mu s^{-1}.
\end{equation}
and experimental values for the decay rate of orthopositronium
{}~\cite{pg21}:
\begin{eqnarray}
\cases{\Gamma^{exp}_{3}(o-Ps)=7.0514(14)\,\mu s^{-1}, \ \ \ \mbox{ref.}~[19a] &
$ $\cr
\Gamma^{exp}_{3}(o-Ps)=7.0482(16)\,\mu s^{-1}, \ \ \ \mbox{ref.} ~[19b]. & $ $
}
\end{eqnarray}
What is the origen of $\sim 10$ standard deviation from predicted theoretical
decay rate?\footnote{M. Samuel, ref.~\cite{sam}, recently proposed the way of
dissolving this problem by using some estimations of next orders. However,
calculations of corrections $\sim \alpha^8$ are not yet carried out. In the
connection with the above three points let me propose another
 way of  estimations of higher orders of $\Gamma^{theor}(o-Ps)$. To my
knowledge nobody
checked  the influence of existence of  axial-vector particle on the
theoretical result.  Moreover, even  in the absence of this particle it is well
known that the Bethe-Salpeter  kernel can
be expanded in $\gamma$-matrix algebra as $I^1 \otimes I^2\oplus
\gamma_5^1\otimes \gamma_5^2 \oplus \gamma_\mu^1 \otimes\gamma_\mu^2 \oplus
(\gamma_\mu\gamma_5)^1 \otimes (\gamma_\mu\gamma_5)^2 \oplus \sigma_{\mu\nu}^1
\otimes \sigma_{\mu\nu}^2$. Calculating the additional  diagrams with   scalar,
pseudoscalar,  axial-vector and tensor exchanges  in the Born approximation and
comparing the contributions of these diagrams with the known result  of  lower
orders on $\alpha$ (in order to define weight of each of diagrams) we could be
able to "catch"  corrections of the higher orders to this quantity, which are
only found from  many-photon  diagrams when usual calculations.}.

The questions induced me and others
to look at the different descriptions of high-spin particles.

\setcounter{equation}{0}
\section{Model}

In the beginning of the sixties Joos~\cite{Joos}, Weinberg~\cite{Weinberg}
and Weaver, Hammer, and Good~\cite{Weaver} have developed
a description of free particles with an arbitrary spin $S=0,{1 \over 2}, 1,{3
\over 2}
\cdots$
The wave functions (WF's) in this approach form the basis
 of the $(S,0)\oplus(0,S)$ representation of the Lorentz group
and are
presented by the $2(2S+1)$-- component spinor:
\begin{equation}\label{eq:psi}
\Psi=\left (\matrix{
\Phi_\sigma\cr
\Xi_\sigma\cr
}\right ).
\end{equation}
The transformation rules
\begin{eqnarray}\label{eq:tran}
\cases{\Phi_{\sigma}(\vec p)=exp\left (+\theta \hat{\vec p}
\hat{\vec J}\right )
\Phi_{\sigma}\left (0 \right ), & $ $\cr
\Xi_{\sigma}\left (\vec p\right )=exp\left (-\theta\hat{\vec p}
\hat {\vec
J}\right ) \Xi_{\sigma}\left (0 \right ) & $ $}
\end{eqnarray}
( $\theta$ is the boost parameter, $tanh\ \theta=
\frac{\mid\vec p\mid}{E}$, $\hat{\vec p}=
\frac{\vec p}{\mid \vec p \mid}$, $\vec p$ is the three-momentum of the
particle,
 $\hat{\vec J}$ is the
angular momentum operator) represent generalization of the
well--known Lorentz boosts for a Dirac particle.
This way of description is on an equal footing to description of  Dirac
particles which
have wave function transformed on the $({1\over 2},0)\oplus  (0, {1\over 2})$
representation.

$2(2S+1)$- component  bispinors in momentum space transform  according  to
\begin{equation}\label{eq:bis}
U(\vec p)={1\over\sqrt{2}}\left (\matrix{
D^S \left (\alpha(\vec p)\right )\xi_\sigma\cr
D^S \left (\alpha^{-1\,+}(\vec p)\right )\xi_\sigma\cr
}\right ),
\end{equation}
for  positive-energy states; and
\begin{equation}
V(\vec p)={1\over \sqrt{2}}\left (\matrix{
D^S \left (\alpha(\vec p)C^{-1}\right )\xi^*_\sigma\cr
D^S \left ( \alpha^{-1\,+}(\vec p)C^{-1}\right )(-1)^{2S}\xi^*_\sigma\cr
}\right ),
\end{equation}
for  negative-energy states with the following notations:
\begin{equation}
\alpha(\vec p)=\frac{p_0+M+(\vec\sigma\vec p)}{\sqrt{2M(p_0+M)}},\quad
C=-i\sigma_2
\end{equation}
being used. $D^{(S)}[\Lambda]$ answers for $(S,0)$ representation of Lorentz
group.

For
particle with spin $S$ the equation has the following form:
\begin{equation}
\left [ \gamma^{\mu_1 \mu_2 \ldots
\mu_{2S}}\partial_{\mu_1}\partial_{\mu_2}\ldots\partial_{\mu_{2S}}+M^{2S}\right
]\Psi(x)=0,
\end{equation}
where $\gamma$- matrices are covariantly defined $2(2S+1) \otimes 2(2S+1)$-
matrices discussed in ref.~\cite{barut} for the first time:
\begin{center}
$\gamma^{\mu_1 \mu_2 \ldots \mu_{2S}}\equiv-i^{2S} \pmatrix{
0 &t^{\mu_1 \mu_2 \ldots \mu_{2S}} \cr
\bar t^{\mu_1 \mu_2 \ldots \mu_{2S}}& 0 \cr
}$\\
\vspace*{4mm}
$\gamma_{5}\equiv\pmatrix{
1 & 0 \cr
0 & -1 \cr
}$,\\
\end{center}
with
\begin{equation}
\bar t^{\mu_1 \mu_2 \ldots \mu_{2S}}=\pm t^{\mu_1 \mu_2 \ldots \mu_{2S}},
\end{equation}
sign being "$-$" when  $\mu$'s contain altogether an odd number of space-like
indices;
\begin{equation}
(\bar t^{\mu_1 \mu_2 \ldots \mu_{2S}})^*=C t^{\mu_1 \mu_2 \ldots
\mu_{2S}}C^{-1};
\end{equation}
and tensor t is defined as following:\\
1) $t^{\mu_1 \mu_2 \ldots \mu_{2S}}_{\sigma\sigma^{\prime}}$ is the $(2S+1)$
matrix
with $\sigma,\sigma^{\prime}=j, j-1,\ldots, -j; \mu_1 \ldots
\mu_{2S}=0,1,2,3$;\\
2) $t$ is symmetric in all $\mu$'s;\\
3)  $t$ is traceless in all $\mu$'s, i. e.,
$g_{\mu_1 \mu_2} t_{\sigma \sigma^{\prime}}^{\mu_1 \mu_2 \ldots \mu_{2S}}=0$;\\
4)  $t$ is tensor under Lorentz transformations,
\begin{eqnarray}
D^{(S)}[\Lambda] t^{\ \mu_1 \mu_2 \ldots \mu_{2S}}
D^{(S)}[\Lambda]^{+}&=&\Lambda_{\nu_1}^{\mu_1}\Lambda_{\nu_2}^{\mu_2}\ldots
\Lambda_{\nu_{2S}}^{\mu_{2S}} t^{\ \nu_1 \nu_2 \ldots \nu_{2S}}\nonumber\\
\bar D^{(S)}[\Lambda] \bar t^{\ \mu_1 \mu_2 \ldots \mu_{2S}} \bar
D^{(S)}[\Lambda]^{+}&=&\Lambda_{\nu_1}^{\mu_1}\Lambda_{\nu_2}^{\mu_2}\ldots
\Lambda_{\nu_{2S}}^{\mu_{2S}}\bar t^{\ \nu_1 \nu_2 \ldots \nu_{2S}}.
\end{eqnarray}
Here
\begin{equation}
\bar D^{(S)}[\Lambda]\equiv D^{(S)}[\Lambda^{-1}]^{+}
\end{equation}
is a matrix corresponding to Lorentz  transformation $\Lambda$ in $(0, S)$
representation.  This way of description is in accordance with Weinberg theorem
as opposed to description on the basis of $A_\mu$,  vector potential. Weinberg
theorem says:\\

\vspace*{1mm}
{\it  The field constructed from massless particle operator $a(\vec p,
\lambda)$ of definite helicity transforms according to representation $(A, B)$
such that $B-A=\lambda$.

E. g., a left-circularly polarized photon with $\lambda=-1$ can be associated
with $(1,0)$,  $({3\over 2}, {1\over 2})$,  $(2,1)$ fields, but not with the
vector potential, $({1\over 2},{1\over 2})$.

It is not the case of massive particle. The field can be constructed out of
$2S+1$ operators $a(\vec p, \sigma)$ for any representation $(A, B)$ that
contains $S$ such that $S=A+B, A+B-1\ldots \mid A-B\mid$, e. g., out of the
vector potential.}\\

\vspace*{1mm}
\noindent
If we use the vector potential this
means that we don't have  well-defined creation and annihilation operators in
the beginning of quantization procedure.

It is also possible to develop the Hamiltonian approach. In the case of
spin-1/2 particle we come to  usual Dirac Hamiltonian
\begin{equation}
{\cal H}=(\vec \alpha\vec p)+m\beta.
\end{equation}
In the case of $S=1$ the Hamiltonian is much more
complicated~\cite{Weaver,mat}:
\begin{equation}
{\cal H}=\frac{2E^2}{2E^2-M^2}(\vec\alpha\vec p)+\beta \left
[E-\frac{2E}{2E^2-M^2}(\vec\alpha\vec p)^2\right ],
\end{equation}
where
\begin{center}
$\vec\alpha\equiv\pmatrix{
\vec S & 0 \cr
0 & -\vec S \cr
}$,\hspace*{5mm}
$\beta\equiv\pmatrix{
0 & 1 \cr
1 & 0\cr
}$,\\
\end{center}

In the papers of P. M. Mathews, e. g., ref.~\cite{mat}, and D. Williams {\it et
al.}, ref.~\cite{will},  Hamiltonian for any spin in this formalism has been
constructed. It has the following form:
\begin{equation}
{\cal H}=\sum_{\nu}{\frac{(E+p)^{4\nu}-M^{4\nu}}{(E+p)^{4\nu}+M^{4\nu}} E
C_{\nu}}+\beta\sum_{\nu}{\frac{2EM^{2\nu}(E+p)^{2\nu}}{(E+p)^{4\nu}+M^{4\nu}}
B_\nu},
\end{equation}
where
\begin{eqnarray}
B_\nu&=&\Lambda_\nu+ \Lambda_{-\nu},\\
C_\nu&=&=\Lambda_\nu- \Lambda_{-\nu}.
\end{eqnarray}
$\Lambda$ is a projection operator to an eigenvalue $\nu$ of helicity operator
$\lambda_p= (\vec \alpha\vec p)/\vert \vec p \vert$, $\nu=-S, -S+1\ldots S$.
The coefficients
$B$ and $C$ are presented in {\it Table II}.
\vspace*{1mm}
\begin{center}
{\it Table II}. Coefficients $B_\mu, C_\mu$ in the Hamiltonian.\\

\vspace*{2mm}
\begin{tabular}{|l|l|}
\hline
&\\
$S$ integer& $S$ half-odd integer\\
&\\
\hline
&\\
$B_0=\prod_{_{\mu=1}}^{^{S}}{\frac{\lambda_p^2-\mu^2}{-\mu^2}}$&\\
&\\
\hline
&\\
$B_{\nu}=B_{-\nu}=\frac{\lambda_p^2}{\nu^2}\prod_{_{\mu=1}}^{\prime^{S}}
{\frac{\lambda_p^2-\mu^2}{\nu^2-\mu^2}}$&$B_\nu
=B_{-\nu}=\prod_{_{\mu=1/2}}^{\prime^{S}}{\frac{\lambda_p^2-\mu^2}{\nu^2-\mu^2}}$\\
&\\
\hline
&\\
$C_\nu=-C_{-\nu}=\frac{\lambda_p}{\nu}\prod_{_{\mu=1}}^{\prime^{S}}
{\frac{\lambda_p^2-\mu^2}{\nu^2-\mu^2}}$&$C_\nu=-C_{-\nu}
=\frac{\lambda_p}{\nu}\prod_{_{\mu=1/2}}^{\prime^{S}}
{\frac{\lambda_p^2-\mu^2}{\nu^2-\mu^2}}$\\
&\\
\hline
\end{tabular}\\
\end{center}

\vspace*{1mm}
\noindent
The prime on the product sign indicates that $\mu=\nu$ is to be excluded.

 Following this description, the spin-one
case~\cite{Sankaranarayanan}-\cite{Tucker} as well as the spin-${3 \over 2}$
case~\cite{Shay} have been investigated in details.  In the case of spin $S=1$
one has
 \begin{equation}
D^{(1,0)}\left (\alpha(\vec p)\right ) = 1+\frac{(\vec S\vec p)}{M}+\frac{(\vec
S\vec p)^2}{M(p_0+M)},
\end{equation}
function appearing in (\ref{eq:bis}); normalization factor is equal to $1$.

In the articles~\cite{Good,Tucker,pursey}  the Feynman diagram technique is
discussed
for  vector particles in a slightly different version of six-component
formalism for
quantum electrodynamics (QED). The following Lagrangian:
\begin{eqnarray}\label{eq:lagtuk}
\lefteqn{{\cal L}^{QED}=\tilde\pi_\mu\bar{\Psi}(x)\Gamma_{\mu\nu}
\pi_{\nu}\Psi(x)-2M^2\bar{\Psi}(x)\Psi(x)+{1 \over
4}F_{\mu\nu}F_{\mu\nu}-}\nonumber\\
&-&\frac{e\lambda}{12}F_{\mu\nu}\bar{\Psi}(x)\gamma_{5,\mu\nu}\Psi(x)-
\frac{e\kappa}{12 M^2}\partial_{\alpha}F_{\mu\nu}\left [\bar{\Psi} (x)
\gamma_{6,\mu\nu,\alpha\beta}\pi_{\beta}\Psi(x)-\right.\nonumber\\
&-&\left.\tilde\pi_\beta\bar \Psi (x)\gamma_{6,\mu\nu,\alpha\beta}\Psi
(x)\right ]
\end{eqnarray}
has been  used there.
In the above formula we have $\pi_{\mu}=-i\partial_{\mu}-eA_{\mu}$,
$\tilde\pi_{\mu}=-i\partial_{\mu}+eA_{\mu}$, $\mu,\nu=1,\ldots 4$;
$F_{\mu\nu}=\partial_{\mu}A_{\nu}-\partial_{\nu}A_{\mu}$ is the
 electromagnetic field tensor; $A_{\mu}$ is the 4- vector of the
electromagnetic field; $\bar{\Psi}, \Psi$  are the six-component  WF of the
massive vector particle.
The following expression has been obtained
for the interaction vertex of a vector particle with a
photon:
\begin{equation}
-e\Gamma_{\alpha\beta}(p+k)_{\beta}-{ie\lambda \over
6}\gamma_{5,\alpha\beta}q_{\beta}+{e\kappa \over
6M^2}\gamma_{6,\alpha\beta,\mu\nu}q_{\beta}q_{\mu}(p+k)_{\nu},
\label{22}
\end{equation}
where  $e$ is  electron charge, $\lambda$ and $\kappa$ are the quantities which
correspond to the magnetic dipole moment and the electric quadrupole moment,
respectively; $M$ is the mass of vector particle;
$\Gamma_{\alpha\beta}=\gamma_{\alpha\beta}+\delta_{\alpha\beta}$;
$\gamma_{\alpha\beta}$ are defined by the formulae
\begin{equation}\label{eq:gam}
\gamma_{ij}=\pmatrix{
0 & \delta_{ij}-S_i S_j- S_j S_i \cr
\delta_{ij}-S_i S_j- S_j S_i & 0 \cr
},
\end{equation}
\begin{equation}
\gamma_{i4}=\gamma_{4i}=\pmatrix{
0 & iS_i \cr
-iS_i & 0 \cr
},\ \ \ \ \ \gamma_{44}=\pmatrix{
0 & 1 \cr
1 & 0\cr
};
\end{equation}
($S_i$ are the  spin matrices for a vector particle).
 $\gamma_{5, \alpha\beta}$; $\gamma_{6,\alpha\beta,
\mu\nu}$ are also $6\otimes 6$- matrices:
\begin{eqnarray}
\gamma_{5,\alpha\beta}&=&i [\gamma_{\alpha\mu}, \gamma_{\beta\mu}],\\
\gamma_{6,\alpha\beta,\mu\nu}&=&
[\gamma_{\alpha\mu},\gamma_{\beta\nu}]_{+}
+2\delta_{\alpha\mu}\delta_{\beta\nu}-[\gamma_{\beta\mu},
\gamma_{\alpha\nu}]_{+}-2\delta_{\beta\mu}\delta_{\alpha\nu}.
\end{eqnarray}
 36 independent Hermitian matrices: ($1, \gamma_5, \gamma_{\alpha\beta},
i\gamma_5 \gamma_{\alpha\beta},
\gamma_{5,\alpha\beta}, \gamma_{6,\alpha\beta,\mu\nu}$) form a complete set of
$6\otimes 6$ matrices.

The propagator of vector meson  in this formulation is
\begin{equation}
G_F(p)=-\frac{i \left [ 2M^2-p_\mu p_\nu
(\gamma_{\mu\nu}-\delta_{\mu\nu})\right ]}{4M^2 (p^2+M^2-i\epsilon)}.
\end{equation}

It was noted in
ref. [5b, p.888] that the equation for 6- component bispinor
\begin{equation}
(\gamma_{\mu\nu}p_\mu p_\nu + M^2)\Psi^{(S=1)}(x)=0
\end{equation}
can be transformed to the equations for left-- and right--circularly
polarized radiation when massless $S=1$ field are  considered.
In fact, if propose that all ways of description of massless vector particles
are
equivalent (at least when interaction is absent and intrinsic momenta are equal
to zero), it is possible following for Weinberg to assume interpretation of
6-component bispinors as
\begin{eqnarray}\label{eq:maj}
\cases{
\Phi=\quad\vec E+i\vec H,& $ $\cr
\Xi=-\vec E+i\vec H& $ $}
\end{eqnarray}
($\vec E$ and $\vec H$ are Pauli vectors). In fact, this is the formulation a
l\'a Majorana~\cite{Recami2}-\cite{Gianetto}.
Thus, we come to the Maxwell's free-space equations
(Eqs. (4.21) and
(4.22) of ref. [5b]):
\begin{eqnarray}\label{eq:Maxwell}
\cases{\vec\nabla\times[\vec E-i\vec H]+i(\partial/\partial t)[\vec E-i\vec
H]=0,& $ $\cr
\vec\nabla\times[\vec E+i\vec H]-i(\partial/\partial t)[\vec E
+i\vec H]=0,& $ $}
\end{eqnarray}
In the following Section we analyze this fact.

As mentioned in {\it Introduction},  Weinberg
equations have some  shortcomings which make it difficult to describe
consistently
high-spin hadron phenomenology.  Recently, considerable development of this
formalism has been undertaken by  D. V.  Ahluwalia {\it et al.}
{}~\cite{ahl,ahl2}. First of all, new equation has been proposed in ref.[7a]
after  analysis of transformation properties of left- and right-
spinors:
\begin{equation}\label{eq:ah}
\left [\gamma_{\mu\nu\ldots \lambda}p^\mu p^\nu\ldots p^\lambda-\wp_{u,v}M^{2S}
I\right ]\Psi (\vec p)=0.
\end{equation}
The incentives of implementing the additional factor $\wp_{u,v}$ are founded by
different intrinsic spin parity of $u$ and $v$ spinors and general structure of
$(S,0)\oplus (0,S)$
theory with the transformation laws (\ref{eq:tran}). As opposed to the
statement of Ryder
{}~\cite[p.44]{Ryder}, the rest spinors satisfy the following equation:
$\phi_R(0)=\wp_{u,v} \phi_L(0)$, $R$ and $L$ mark left- and right- spinors. In
ref. [7a,b] it was found that this
additional factor keeps for bosons (this is not the case for fermions) in the
configurational-space
equation, e. g., for spin $S=1$ one has
\begin{equation}
\left [\gamma_{\mu\nu}\partial^\mu \partial^{\nu}- \wp_{u,v}M^{2} I_6\right
]=0,
\end{equation}
$I_6$ is the unit $6\otimes 6$- matrix.
Thus, $(S,0)\oplus (0,S)$ representation for massive particles is a realization
of the Bargmann-
Wightman-Wigner-type quantum field theory~\cite{Bargm}, in which a boson
(described by $u$ spinors) and its antiboson (described by $v$ spinors) have
opposite  intrinsic parities.  Let us mention that in the case of $({1\over 2},
{1\over 2})$
representation on which vector potential transforms particle-antiparticle pairs
have the same intrinsic parity. In ref. [7d] it was pointed out that Prof. C.
Rubbia suggested to use the process $\gamma\gamma\rightarrow W^+ W^-$ in order
to check if $W^+$ and $W^-$ carry the same or opposite intrinsic parity.

However, even with this equation (\ref{eq:ah}) one have tachyonic solutions
$E^2=\vec p^2-M^2$ which are to be explained.  In ref. [7a]  interpretation on
the basis of  some reminiscences of $m^2 <0$ of the simplest versions of
quantum field theories with spontaneous symmetry breaking has been proposed.
Namely, introduction quartic interaction
$\lambda (\bar\Psi(x) \Psi (x))^2$ could lead to the breaking of rotational
symmetry (cosmic vortices  are possible manifestation of this breaking) and
appearance of   quartet of particles $\left \{ {\cal N}^{\pm}, {\cal N}^0,
\hat\eta \right \}$, with one of them  $\hat\eta$ being a Goldstone-like
spin-one massless particle.

\setcounter{equation}{0}
\section{Main Results}

The main results obtained by the author in the preceding papers~\cite{Dvoeglaz}
are:

1) The scalar Lagrangian of Weinberg's theory (the case of  massless $S=1$
particles and a l\'a Majorana interpretation of Weinberg's WF, Eq.
(\ref{eq:maj}))
\begin{equation}\label{eq:la}
{\cal L}^{JW}=\partial_\mu\bar\Psi\gamma_{\mu\nu}\partial_\nu\Psi
\end{equation}
is shown to be equivalent to the Lagrangian of free massless
skew-symmetric field ${\cal L}^H=F_k F_k/8$
($F_k=i\epsilon_{kjmn}F_{jm,n}$), presented by
Hayashi (1973), ref.~\cite{Hayashi}.
It can be rewritten
\begin{eqnarray}\label{eq:Lagran}
{\cal L}^{JW}&=&(\partial_\mu F_{\nu\alpha})(\partial_\mu
F_{\nu\alpha}) -
2(\partial_\mu F_{\mu\alpha})(\partial_\nu F_{\nu\alpha})
+ 2(\partial_\mu
F_{\nu\alpha})(\partial_\nu F_{\alpha\mu})=\nonumber\\
&=&-4{\cal L}^{H}-2(\partial_{\mu}
F_{\mu\alpha})(\partial_{\nu} F_{\nu\alpha}),
\end{eqnarray}
what confirms the above statement, taking into account
the possibility of the use of
the Fermi method {\it mutatis mutandis} as in ref.~\cite{Hayashi}.
The second
term in (\ref{eq:Lagran}) can be excluded by means of generalized
Lorentz
condition (which is just well-known Maxwell  motion
equation)\footnote{Let us
mention some analogy with the potential formulation of QED.
In some sense the
Lagrangian (\ref{eq:Lagran}) corresponds to the choice of
"gauge-fixing" parameter $\xi=-1$, ${\cal L}^{H}$  of
ref.~\cite[formula
(5)]{Hayashi} corresponds to the "Landau gauge" ($\xi =0$),
and ${\cal L}^{H}$
(formula (9) of the cited paper),  to the "Feynman
gauge"($\xi=1$).}.
The Lagrangian describes massless particles with the longitudinal
physical components only. The transversal components are removed by means
of the new "gauge" transformation:
\begin{equation}\label{eq:gauge}
F_{\mu\nu}\rightarrow
F_{\mu\nu}+A_{[\mu\nu]}=F_{\mu\nu}+\partial_{\nu}
\Lambda_{\mu}-
\partial_{\mu}\Lambda_{\nu}.
\end{equation}

2) The vector Lagrangian, proposed in ref. [34b],
\begin{equation}\label{eq:Lagr}
{\cal L}_{\alpha}=-i\bar\Psi\gamma_{\alpha\beta}\partial_{\beta}\Psi
+i(\partial_{\beta}\bar\Psi)\gamma_{\alpha\beta}\Psi
\end{equation}
gives the dynamical invariants which are equivalent to the ones found by Lipkin
(1964)
and Sudbery (1986), ref.~\cite{Lipkin}. The tensor energy-momentum  has the
following components:
\begin{eqnarray}
T_{\{i}^{\quad 4\}4}&=&(\vec E\vec\nabla)\vec H-(\vec H\vec\nabla)
\vec E+
\vec E(\nabla\vec H)-\vec H(\vec\nabla\vec E),\\
T_{\{4}^{\quad 4\}4}&=&\vec E [\vec\nabla\times\vec H]-\vec
H[\vec\nabla\times\vec E],\\
T_{\{i}^{\quad j\}4}&=&\vec\nabla\vee
\left [\vec E\times\vec H\right ],\\
T_{[i}^{\quad 4]4}&=&-i[(\vec E\vec\nabla)\vec E+
(\vec H\vec\nabla)\vec H+
\vec E(\vec\nabla\vec E)+\vec H(\vec\nabla\vec H)],\\
\tilde T_{i}\hspace*{7mm}&=&{1\over 2}\epsilon_{ijk}
T_{[j}^{\quad k]4}=\left [(\vec
E\vec\nabla)\vec H-
(\vec H\vec\nabla)\vec E+\vec H(\vec\nabla\vec E)-
\vec E(\vec\nabla\vec
H)\right ];
\end{eqnarray}
and for spin tensor, as opposed to ref. [36b], we obtained
\begin{equation}
S^{4,ij}_{4}=0,
\end{equation}
but
\begin{equation}
S^{4,4i}_{4}=-4\left [\vec E\times\vec H\right ]_i.
\end{equation}
At last, we have the same expressions for $J^{\mu}_{\alpha}$
as in ref. [36b]:
\begin{eqnarray}
J_{44}&=&-2(\vec E^2+\vec H^2),\\
J_{4i}&=&4i\epsilon_{ijk}E_j H_k,\\
J_{ij}&=&2[(\vec E^2+\vec H^2)\delta_{ij}-E_i E_j - H_i H_j],
\end{eqnarray}
which are the components of energy-momentum tensor in
the common-used
formulation of QED.
Thus, "charge" is identified with the energy
density of the
field and  energy-momentum conservation is associated not with translational
invariance
but with invariance under duality rotations;

3)  Since the result of  item (1)  is in contradiction with Weinberg theorem
about connection
between $(A,B)$- representation of the Lorentz group and helicity of particle
($B-A=\lambda$, see Section 2) and, moreover, the Weinberg massless equations
[5b] admit the acausal ($E\neq\pm p$) solutions [6c],
new interpretation of the  Weinberg's $S=1$ bispinor has been proposed.
It is based on  the use of  the axial-vector potential $\tilde A_k$ which is
constructed from the  strength tensor
$\tilde F_{\alpha\beta}={i \over 2}\epsilon_{\alpha\beta\mu\nu} F_{\mu\nu}$,
dual to the electromagnetic field tensor,
\begin{eqnarray}\label{eq:pse}
\cases{\Phi_{k} = \tilde{A}_{k} + iA_{k},& $ $\cr
\Xi_{k} = \tilde{A}_{k} - iA_{k}.& $ $}
\end{eqnarray}

4) The interaction Hamiltonian (two $S=1/2$ particles and one  massless $S=1$
particle), which was discussed, e. g., by  Marinov (1968), ref.~\cite{Marinov},
\begin{eqnarray}\label{eq:Ham}
{\cal H}_{\Psi\psi\psi}=g\sum_{\mu_1\,\mu_2\,\mu_3}\left (\matrix{
S_1 & S_2 & S_3 \cr
\mu_1 & \mu_2 & \mu_3 \cr
}\right )
\phi^{\mu_1}_{(S_1)}\phi^{\mu_2}_{(S_2)}\phi^{\mu_3}_{(S_3)}\pm\nonumber\\
\pm \left (\matrix{
S_1 & S_2 & S_3 \cr
\dot{\mu}_1 &\dot{\mu}_2 & \dot{\mu}_3 \cr
}\right )
\chi^{\dot{\mu}_1}_{(S_1)}\chi^{\dot{\mu}_2}_{(S_2)}\chi^{\dot{\mu}_3}_{(S_3)},
\end{eqnarray}
\\
where
$\left (\matrix{
S_1 & S_2 & S_3 \cr
\mu_1 & \mu_2 & \mu_3 \cr
}\right )$
are the Wigner $3j$- symbols, appears to lead to the equations
\begin{equation}\label{eq:Dir}
i\hbar\frac{\partial\psi}{\partial t}=c\vec\alpha\cdot(\vec p-e\vec
A-ie\gamma_5\vec{\tilde{A}})\psi+mc^2\beta\psi,
\end{equation}
which could give the equation very similar to  the Dirac oscillator
ones\footnote{ In the version of Moshinsky~\cite{Moshinsky} interaction has
been introduced in Dirac equation in the form: \ $(\vec\alpha\vec r)\beta$; in
the scalar version~\cite{Dixit}  interaction  has been  introduced as
$m\rightarrow m\left [1+{\omega \over c} r\gamma_5\right ]$.}:
\begin{eqnarray}\label{eq:osc}
\cases{(E^2-m^2)\xi=\left [\vec p\,^2+m^2\omega^2 r^2+2iEm\omega(\vec\sigma\vec
r)+3m\omega+4m\omega \vec S[\vec r \times \vec p]\right ]\xi,& $ $\cr
(E^2-m^2)\eta=\left [\vec p\,^2+m^2\omega^2 r^2+2iEm\omega(\vec\sigma\vec
r)-3m\omega-4m\omega \vec S[\vec r \times \vec p]\right ]\eta.& $ $}
\end{eqnarray}
Similarly to~\cite{Dixit} these equations include the term $\sim(\vec\sigma\vec
r)$ and are not
invariant under parity transformation. To keep parity conservation it is
necessary to assume that $\omega$, a
frequency, is a pseudoscalar quantity, what means complicated dispersion law.
However, similarly to the case of interaction of four bispinors,
''irregular'' (terminology of Marinov) invariants (where upper and down
components of bispinors are mixed) for interaction between such types of fields
were pointed out in ref.~\cite{Marinov} to be possible (see also ref. [5b,
p.890]).
In fact, in some sense the particle with spin one "consists" of two particles
with spin $1/2$ since vector and  axial-vector have the  transformation  laws
which are analogous to the ones of some combination of dotted and undotted
spinor, ref.~\cite{berest}:
\begin{eqnarray}
\xi^{\alpha\dot\beta} (vector)\hspace*{7mm}&\sim&\xi^\alpha
H^{\dot\beta}+\Xi^\alpha \eta^{\dot\beta}\\
\xi^{\alpha\dot\beta} (axial-vector)&\sim&\xi^\alpha H^{\dot\beta}-\Xi^\alpha
\eta^{\dot\beta}
\end{eqnarray}
This fact gives the opportunity to construct
"irregular" invariant which leads to the same Dirac oscillator equations as
proposed in ref.~\cite{Moshinsky}.

The appearance of the new term ($2iEm\omega(\vec\sigma\vec r)$)  can also be
explained
by the fact that it is possible to add in the formula (5) of the
paper~\cite{Moshinsky} both the term $-im\omega \beta\vec r$, which corresponds
to the addition $\alpha_i\wedge\alpha_4 R_{4i}$ (where $R_{4j}=ir_j$), and the
one ${m\omega\over 2} [\vec\alpha\times~\vec r]$, which corresponds to the
interaction term ${1\over 2}
\alpha_i\wedge\alpha_j R_{ij}$ (where $R_{ij}=\epsilon_{ijk}r_k$), in
accordance with bivector construction rules as the expansion  in Clifford
algebra in the Minkowsky 4- dimensional space~\cite{Jancewicz}.
Thus, the interaction term for the Dirac oscillator is possible to define:
\begin{equation}
R=-m\omega\alpha_i\wedge\alpha_4 R_{4i}+\frac{m\omega}{2}\alpha_i\wedge\alpha_j
R_{ij}.
\end{equation}
(cf. formula (32) in ref.~\cite[p. 244]{Jancewicz}). Thus, instead of the
minimal form of electromagnetic interaction ($\gamma_\mu
A_\mu$) we have the bivector form interaction (similarly to  introduction of
the Pauli term but  not applying to antisymmetric field tensor).

We have also deduced  the Hamiltonians for interaction of  various spin
particles from Eq. (\ref{eq:Ham}) following ideas of decomposing left- and
right- spinors into scalar (vector)
and pseudoscalar (axial-vector) parts. E. g., the Hamiltonian for interaction
two spin-$1/2$ particles with spin- $0$ particle ($\Phi^{(S=0)}=\phi +
i\tilde\phi$ and $\Xi^{(S=0)}=\phi - i\tilde\phi$) is
\begin{equation}
{\cal H}_{\bar\psi\psi\phi}= \frac{g}{\sqrt{2}}\left \{\bar\psi \psi \phi
-i\bar \psi \gamma_5 \psi \tilde \phi \right \}.
\end{equation}

5) The presented formalism has been used for calculation of the scattering
amplitude for two gluon interaction
\begin{eqnarray}
\lefteqn{\hat T^{(2)}_{gg} (\vec k(-)\vec p, \vec p) = -3g_V^2 \left\{
\frac{\left
[p_0
(\Delta_0 +M) + (\vec p
\vec\Delta)\right ]^2 -M^3 (\Delta_0+M)} {M^3 (\Delta_0 -M)}
+\right.}\nonumber\\
&+&\left.\frac{i (\vec S_1+\vec S_2)\left [\vec p\vec \Delta\right ]} {\Delta_0
-M}
\left [ \frac{p_0 (\Delta_0 +M)+\vec p\vec\Delta}{M^3}
\right ] + \frac{(\vec S_1\vec\Delta)(\vec S_2\vec\Delta)-(\vec
S_1\vec S_2)\vec\Delta^2}{2M(\Delta_0-M)}-\right.\nonumber\\
&-&\left.\frac{1}{M^3}\frac{\vec S_1\left [\vec p \vec \Delta\right ]\vec S_2
\left [\vec
p\vec\Delta\right ]}{\Delta_0-M}\right\}. \label{212}
\end{eqnarray}
Analogously to the earlier works devoted to fermion-fermion
interaction~\cite{a11}, we have
\begin{equation}
\vec\Delta \equiv \Lambda^{-1}_{\vec p} {\vec k} \equiv \vec k(-) \vec p = \vec
k-\frac{\vec p}{M} (k_0 - \frac{\vec k\vec p}{p_0 +M}),
\end{equation}
\begin{equation}
\Delta_0 \equiv (\Lambda^{-1}_{\vec p} k)_0 = (k_0 p_0 -\vec k\vec p)/M,
\end{equation}
which are the 4-vector of momentum transfer in the Lobachevsky space.
The remarkable fact is that the amplitude coincides with the amplitude for
interaction   of two spinor particles   in the Lobachevsky space  ($p_0^2-\vec
p\,^2=M^2$) except for obvious substitutions $1/[2m(\Delta_0 - m)]\rightarrow
1/\vec \Delta^2$ and $\vec S\rightarrow \vec\sigma$.
Furthermore,  three - dimensional covariant equal-time (quasipotential)
equations have been found for the composite systems, namely, fermion -- boson
of $S=1$ (quark-diquark system), ref. [34d], and two bosons of spin $S=1$
(gluonium), ref. [34g] and  relativistic partial-wave quasipotential equations
have been obtained for the singlet gluonium state, the triplet one and the
5-plet gluonium state in  the relativistic configurational representation
(RCR), ref. [34e], which is  generalization of  $x$-representation. Shapiro
plane-wave functions~\cite{Shap} are to be used instead of Fourier
transformation, i. e.
\begin{equation}
V(r,\vec n; \vec p) = {1\over (2\pi)^{3}} \int d\Omega_{\Delta}\, \xi^{*} (\vec
\Delta; \vec r) V (\vec \Delta, \vec p), \label{31}
\end{equation}
for the quasipotential and
\begin{equation}
\Psi_{\sigma_1\sigma_2} (r,\vec n) = {1\over (2\pi)^{3}} \int d\Omega_p\, \xi
(\vec p; \vec r) \Psi_{\sigma_1\sigma_2} (\vec p), \label{32}
\end{equation}
for the bound state WF. The integration measure is
\begin{equation}
d\Omega_p \equiv d^3\vec p /\sqrt{1+\vec p^{\,2}/M^2}.
\end{equation}
It is the invariant measure on the hyperboloid, $p_0^2-\vec p^{\,2}=M^2$. The
system of functions  $\xi(\vec{p};\vec{n},r)$ is the complete orthogonal system
of functions in the Lobachevsky space,
\begin{equation}
\xi (\vec p; \vec n; r)\equiv (\frac{p_0-\vec p\vec n}{M})^{-1-irM}.
\end{equation}

6) In the connection with the importance of axial-vector potential let me also
reproduce
the amplitude for interaction between two fermions mediated  axial-vector
massless
particle in the Lobachevsky space.
\begin{eqnarray}
\hat T_{AV}^{(2)}&=& -g_{AV}^2 \left \{ (\vec \sigma_1\vec
\sigma_2)\frac{\Delta_0+m}{\Delta_0-m}+
2(\vec \sigma_1\vec p)(\vec \sigma_2\vec
p)\frac{\Delta_0+m}{m^2(\Delta_0-m)}+\right.\nonumber\\
&+&\left. 2\left [(\vec \sigma_1\vec p)(\vec \sigma_2\vec \Delta)+(\vec
\sigma_1\vec\Delta)(\vec\sigma_2\vec p)\right
]\frac{p_0}{m^2(\Delta_0-m)}+\right.\nonumber\\
&+&\left.  (\vec\sigma_1\vec\Delta)(\vec
\sigma_2\vec\Delta)\frac{2p_0^2-m^2}{m^2\Delta^2}\right \}
\end{eqnarray}
 It can be used for realization of calculation program  mentioned in {\it
footnote 1}.

7) The relativistic analogue of the Shay-Good Hamiltonian has also been
obtained in the Lobachevsky space. The new magnetic momentum
vector has been defined [34c,h].

\begin{eqnarray} \label{eq:ham}
\lefteqn{\hat H=\left\{-\frac{e}{M}\left [(p_{\mu}A_{\mu})\left
(1+\frac{1}{3}\frac{\Delta_{0}-M}{M}\left
(1-\lambda-\frac{2\kappa}{3}\frac{\Delta_{0}-M}{M}\right )\right
)+\right.\right.}\nonumber\\
&+&\left.\left.\frac{1}{2M^2}\left (\vec\Theta \vec E\right )\left
(1+\lambda+\frac{\Delta_{0}-M}{M}\kappa \right )-\frac{1}{2M^2}\left
(\vec{\Xi}\vec B\right )\left (1+\lambda+\frac{\Delta_{0}-M}{M}\kappa\right )+
\right.\right.\nonumber\\
&+&\left.\left.\frac{(p_{\mu}A_{\mu})}{2M(\Delta_{0}+M)}
Q_{ik}\Delta_{i}\Delta_{k}\left (1-\lambda-2\kappa-\frac{2}{3}
\frac{\Delta_{0}-M}{M}\kappa\right )\right ]+\right.\nonumber\\
&+&\left. 2e^{2}\left[ A^{2}_{\mu}\left (
1+\frac{2\kappa}{3}\frac{\Delta_{0}-M}{M}\right )+A^{2}_{\mu}\frac{(\vec
S\vec\Delta)^{2}}{M(\Delta_{0}+M)}\left
(1+\frac{2\kappa}{3}\frac{\Delta_{0}-M}{M}
-\frac{\kappa}{2}\frac{\Delta_{0}
+M}{M}\right )-\right.\right.\nonumber\\
&-&\left.\left.\frac{1}{M^3}(p_{\mu}A_{\mu})(W_{\nu}A_{\nu})(\vec
S\vec\Delta)-\frac{1}{M^2}(W_{\mu}A_{\mu})(W_{\nu}A_{\nu})\left
(1+\frac{\Delta_{0}-M}{M}\kappa\right)-\right.\right.\nonumber\\
&-&\left.\left.\frac{1}{M^2}(W_{\mu}A_{\mu})(W_{\nu}A_{\nu})\frac{(\vec
S\vec\Delta)^2}{M(\Delta_{0}+M)}(1-2\kappa)\right ]\right \}\otimes
D^{(1)}\left\{V^{-1}(\Lambda_{\vec p},\vec k)\right\}
\end{eqnarray}
In  Eq. (\ref{eq:ham}) the vectors
\begin{eqnarray}
\vec\Theta &=&(\Sigma_{(41)},\Sigma_{(42)},\Sigma_{(43)}),\\
\vec\Xi &=& i\left(\Sigma_{(23)},\Sigma_{(31)},\Sigma_{(12)}\right),
\end{eqnarray}
constructed from the tensor components $\Sigma_{(\mu\nu)}(\vec p) $,
\begin{equation}
\Sigma_{(\mu\nu)}(\vec p) = \frac{1}{2}\left \{ W_{\mu}(\vec p)W_{\nu}(\vec p)
-
W_{\nu}(\vec p)W_{\mu}(\vec p)\right\},
\end{equation}
have been used. Here $W_{\mu}(\vec p)$ is the Pauli-Lyuban'sky 4- vector of
relativistic spin,  $Q_{ik} $ is the quadrupole momentum tensor for vector
particle. Thus, the vector $\frac{e}{2M^3}\vec\Xi$ could be  defined as the
vector of the magnetic momentum  for $S=1$ particle  moving with the linear
momentum $\vec p$.

\newpage
\section{Conclusions}

Conclusions are:

--- Searches of  satisfactory theories for description of high-spin  particles
(as well as alternative formulation of vector boson theory) do have definite
reasons because of some shortcomings  of usual models.

---  Weinberg's $2(2S+1)$- formalism, which is used in the present work, is
very similar to the standard Dirac's approach to spinor particles and,
therefore, seems to be convenient for practical calculations.

--- Interpretation of \, WF of  massless $S=1$ particle, which had been  given
by Weinberg [5b],  is not sufficiently satisfactory.

--- We built, in fact, a bivector form of interaction between spinor particle
and Weinberg's vector boson  instead of a minimal form of interaction in the
case of the proposed interpretation of the Weinberg's WF (Eq. \ref{eq:pse}).

--- Estimations of eventual influence of this model on the experimental results
deserve further elaboration.\\

\scriptsize{{\it Note Added.} I am very grateful to Prof. J. S. Dowker for
drawing
my attention to his papers and papers of Morgan and Joseph (see {\it Appendix})
who proposed tensor Lagrangians which the Lagrangian (\ref{eq:Lagr}) is similar
to.}

\newpage
\sloppy
\voffset=-3cm
\hoffset=-2.7cm
\addtolength{\topmargin}{72pt}
\oddsidemargin20mm
\evensidemargin20mm
\addtolength{\textwidth}{20mm}
\setlength{\columnsep}{10mm}
\setlength{\columnseprule}{.07mm}
\pagenumbering{roman}
\setcounter{page}{1}
\footnotesize{
\twocolumn
\section*{{\tt APPENDIX}}

Here we present  a overview of   literature related to
$2(2S+1)$ component  formalism for description of high spin particles.

\section*{1959-70}

\begin{enumerate}


\item
R. H. Good, jr.  Theory of particles with zero rest-mass.
In {\em Lectures in Theoretical Physics. Summer Institute for Theoretical
Physics. 1958. Vol. 1.} \  Eds. W. E.
Brittin and L. G. Dunham. Interscience Publ. Inc.  New York, 1959, p. 30-81

\item
H. Joos.  Zur Darstellungstheore der inhomogenen Lorentzgruppe als
 Grundlage quantenmechanisher Kinematik. \fp {10}{3}, 65-146 (1962)

\item
A. O. Barut, I. Muzinich and D. N. Williams. Construction of invariant
scattering amplitudes for arbitrary spins and analytic continuation in total
angular
momentum. \pr {130}{1}, 442-457 (1963)

\item
S. Weinberg. Feynman rules for any spin. \pr {133}{5B},  1318-1332  (1964)

\item
S. Weinberg. Feynman rules for any spin. II. Massless
particles. \pr {134}{4B}, 882-896 (1964)

\item
D. L. Weaver, C. L. Hammer and R. H. Good, jr. Description of a
particle with arbitrary mass and spin. \pr {135}{1B}, 241-248 (1964)

\item
S. Weinberg. The quantum theory of massless  particles. In  {\em Lectures on
Particles and Field Theory.  Brandeis Summer Institute  in Theoretical Physics.
1964. Vol. 2.} \  Eds. S. Deser and
K. W. Ford. Prentice Hall Inc., New Jersey, 1965, p. 405-485

\item
D. N. Williams. The Dirac algebra for any spin.  In  {\em Lectures in
Theoretical Physics. Summer Institute for Theoretical Physics. 1964. Vol. VII A
- Lorentz Group. }
\ Eds. W.E. Brittin and A. O. Barut. University of Colorado Press, Boulder,
1965,  p. 139-172

\item
R. Shaw.  Unitary representations of the inhomogeneous Lorentz group. \nc
{33}{4}, 1074-1090 (1964)

\item
D. L. Pursey. General theory of covariant particle equations. 1\apny {32}{1},
157-191 (1965)

\item
A. Sankaranarayanan and R. H. Good, jr.  Spin-one equation. \nc {36}{4},
1303-1315  (1965)

\item
R. Shaw.  Zero-mass and nondecomposable representations. \nc {37}{3}, 1086-1099
 (1965)

\item
D. J. Candlin.  Physical operators and representations of the inhomogeneous
Lorentz group. \nc {37}{4}, 1396-1406  (1965)

\item
A. Sankaranarayanan and R. H. Good, jr. Position operators in relativistic
single-particle theories. \pr {140}{2B}, 509- 513 (1965)

\item
D. Shay, H. S. Song and R. H. Good, jr. Spin three-halves wave equations.
\ncs {3}{3}, 455-476  (1965)

\item
A. Sankaranarayanan. Covariant polarization theory of spin-one particles.
\nc {38}{2}, 889-906  (1965)

\item
T. A. Morgan and D. W. Joseph. Tensor Lagrangians and generalized conservation
laws
for free fields. \nc {39}{2}, 494-503 (1965)

\item
A. Sankaranarayanan. Covariant multipole polarization operators and density
matrices for any spin.  \nc {A41}{4}, 532-542 (1966)

\item
P. M. Mathews. Relativistic Schr\"odinger equations for
particles of arbitrary spin. \pr {143}{4}, 978-985 (1966)

\item
P. M. Mathews. Invariant scalar product and observables in a relativistic
theory of particles of arbitrary spin. \pr {143}{4}, 985-989 (1966)

\item
S. A. Williams, J. P. Draayer and T. A. Weber. Spin-matrix polynomial
development
of the Hamiltonian for a free particle of arbitrary spin and mass. \pr
{152}{4}, 1207-1212 (1966)

\item
D. Shay.  Ph. D. thesis 67-2092, Iowa State University, 1966

\item
J. S. Dowker and Y. P. Dowker. Interactions of massless particles of arbitrary
spin.
\prsl {294}{1437}, 175-194  (1966)

\item
J. S. Dowker and Y. P. Dowker. Particles of arbitrary spin in curved
spaces. \pps {87}{1}, 65-78 (1966)

\item
J. S. Dowker. Coulomb scattering for arbitrary spin. \pps {89}{}, 353-364
(1966)

\item
J. S. Dowker. A wave equation for massive particles of arbitrary spin. \prsl
{297}{1450}, 351-364 (1967)

\item
J. S. Dowker. A note on the Rainich problem. \pps {91}{1}, 23-27 (1967)

\item
J. S. Dowker. On the 'Zilch' of gravitational fields. \pps {91}{1}, 28-30
(1967)

\item
J. S. Dowker. Consistent higher-spin wave equations. \ncs {5}{3}, 734-738
(1967)

\item
Wu-Ki Tung. Relativistic wave equations and field theory for arbitrary spin.
\pr {156}{5}, 1385-1398 (1967)

\item
J. S. Dowker. The generalized Bel-Robinson tensor as a generator. \jpa {1}{2},
277-279  (1968)

\item
T. J.  Nelson and R. H. Good,jr.  Second-quantization process for particles
with any spin and with internal symmetry. \rmp {40}{3}, 508-522 (1968)

\item
C. L. Hammer, S. C. McDonald and D. L. Pursey. Wave equations on a hyperplane.
\pr {171}{5}, 1349-1356 (1968)

\item
M. S. Marinov.  Construction of invariant amplitudes for
interactions of  particles with any spin. \apny {49}{3}, 357-392 (1968)

\item
J. S. Dowker and M. Goldstone. The geometry and algebra of the representation
of the Lorentz group. \prsl {303}{1474}, 381-396 (1968)

\item
D. Shay and R. H. Good, jr.  Spin-one particle in an external
electromagnetic field. \pr {179}{5},1410-1417  (1969)

\item
Y. Frishman and C. Itzykson. Massless particles and fields. \pr {180}{5},
1556-1571 (1969)

\item
S. Weinberg. Feynman rules for any spin. III. \pr {181}{5}, 1893-1899 (1969)

\item
P. M. Mathews. Massless particles: relation between helicity and the
transformation character of the wave function. \jmps {4}{1}, 58-63 (1970)

\item
T. J. Nelson and R. H. Good,jr. Lorentz-covariant matrices for
elementary-particle theories as polinomials in the spin matrices.  \jmp
{11}{4}, 1355-1359 (1970)


\noindent
\section*{1971-80}


\item
P. M. Mathews. Wave equations for arbitrary spin.  In  {\em Lectures in
Theoretical Physics. Summer Institute for Theoretical Physics. 1971. Vol. XII C
} Gordon  \& Breach, New York, 1971, p. 139

\item
L. D. Krase, Pao Lu and R. H. Good, jr.  Stationary states of a spin-1 particle
in a constant magnetic field. \pr {D3}{6}, 1275-1279  (1971)

\item
M. Seetharaman, J. Jayaraman and P. M. Mathews. Arbitrary-spin wave equations
and Lorentz invariance. \jmp {12}{5}, 835-840 (1971)

\item
R. H. Tucker and  C. L. Hammer. New quantum electrodynamics for vector mesons.
\pr {D3}{10}, 2448-2460  (1971)

\item
C. L. Hammer and R. H. Tucker. A method of quantization for relativistic
fields. \jmp {12}{7}, 1327-1333  (1971)

\item
M. Seetharaman, J. Jayaraman and P. M. Mathews. Relativistic wave equations:
proper Lorentz invariance and invariance under discrete transformations. \jmp
{12}{8},  1620-1622 (1971)

\item
P. M. Mathews and M. Seetharaman. Arbitrary spin fields: spectral
representation for two-point functions and the connection between spin and
statistics. \np {B31}{2}, 551-569 (1971)

\item
R. N. Faustov. Relativistic transformation of one-particle wave functions (in
Russian). Preprint ITF-71-117P, Kiev, Oct. 1971

\item
Wu-yang Tsai and A. Yildiz. Motion of charged particles in a homogeneous
magnetic field. \pr {D4}{12}, 3643-3648  (1971)

\item
T. Goldman and Wu-yang Tsai. Motion  of charged particles in a homogeneous
magnetic field. II. \pr {D4}{12}, 3648-3651  (1971)

\item
Wu-yang Tsai. Motion of spin- $1$ particles in Homogeneous magnetic field --
Multispinor formalism.  \pr {D4}{12}, 3652-3657  (1971)

\item
C. L. Hammer and  T. A. Weber. Field theory for stable and unstable particles.
\pr {D5}{11}, 3087-3102 (1972)

\item
M. Seetharman and P. M. Mathews. Poincar\'e and TCP invariance in the
determination of wave equations for particles of arbitrary spin. \jmp {13}{7},
938-943 (1972)

\item
J. Jayaraman. Proper Poincar\'e invariance in the determination of
arbitrary-spin
wave equations: constraints from discrete symmetries. \nc {13A} {4}, 877-896
(1973)

\item
J. Jayaraman. Invariant scalar products and quantization of general Poincar\'e-
invariant wave equations. \nc {14A} {2}, 343-362 (1973)

\item
C. L. Hammer and B. deFacio. Unstable Goldstone bosons and the elimination of
scalar particles from gauge theories. \pr {D10}{4}, 1225-1236  (1974)

\item
B. deFacio and C. L. Hammer. Remarks on the Klauder phenomenon. \jmp {15}{7},
1071-1077  (1974)

\item
R. F. Guertin. Relativistic Hamiltonian equations for any spin. \apny {88}{2},
504-533  (1974)

\item
K. Achenbach, Weinberg fields, propagators and van Hove model (in German).
Thesis,
Bonn-IR-75-45, Oct. 1975

\item
Yu. V. Novozhilov. {\em  Introduction to Elementary Particle Theory.} Pergamon
Press,
Oxford, 1975, section 5.3

\item
N. H. Fuchs. How to construct a relativistic $SU(6)$ classification symmetry
group.
\pr {D11}{6}, 1569-1579  (1975)

\item
L. O' Raifeartaigh. Weight diagrams for superfields. \np {B89}{3}, 418-428
(1975)

\item
W. Becker and H. Mitter. Relativistic theory of vector mesons in laser fields.
\apa {43}{3-4},
335-339  (1975)

\item
R. F. Guertin. Foldy-Wouthuysen transformations for any spin. \apny {91}{2},
386-412 (1975)

\item
M. J. King, L. Durand and K. C. Wali. A resonance-sum model for Reggeization in
the scattering of particles with arbitrary spin. \pr {D13}{5}, 1409-1429
(1976)

\item
P. Hoyer and H. B. Thacker. Resonance contributions to finite energy sum rules.
\pr {D13}{7}, 2033-2063  (1976)

\item
H. van Dam and L. C. Biedenharn. The kinematics of a Poincar\'e covariant
object having indecomposable internal structure. \pr {D14}{2}, 405-417 (1976)

\item
R. F. Guertin and C. G. Trahern. Linear relativistic Hamiltonians. \apny
{98}{1}, 50-69 (1976)

\item
R. A. Krajcik and M. M. Nieto. Foldy-Wouthuysen transformations in an
indefinite metric space.
4. Exact, closed form expressions for first order wave equations. \pr {D15}{2},
426-432 (1976)

\item
G. Post. Properties of massless relativistic fields under the conformal group.
\jmp {17}{1}, 24-32  (1976)

\item
R. F. Guertin and T. L. Wilson. Noncausal propagation in spin- $0$ theories
with external field interactions. \pr {D15}{6}, 1518-1531  (1977)

\item
R. F. Guertin and T. L. Wilson. Sakata-Taketani spin $0$ theory with external
field interactions: Lagrangian formalism and causal properties. \apny {104}{2},
427-459  (1977)

\item
G. Grensing. Quantized fields over de Sitter space. \jpa {10}{10}, 1687-1719
(1977)

\item
H. M. Ruck and W. Greiner. A study of the electromagnetic interaction given by
relativistic spin-1 wave equations in elastic scattering of polarized spin-1
nuclei or
mesons. \jpg {3}{5}, 657-680 (1977)

\item
M. D. Birrell. Stress tensor conformal anomaly for Weinberg-type fields in
curved spacetimes.
\jpa{12}{3}, 337-351  (1979)

\item
B. Vijayalakshmi, M. Seetharaman and P. M. Mathews. Consistency of spin 1
theories in
external electromagnetic fields. \jpa {12}{5}, 665-677  (1979)

\item
J. Le\'on, M. Quir\'os and J. Ramirez Mittelbrunn. Group content of the
Foldy-Wouthhuysen transformation and nonrelativistic limit for arbitrary spin.
\jmp {20}{6}, 1068-1076  (1979)

\item
J. O.  Eeg. Calculations of Dirac spinor amplitudes by means of complex Lorentz
transformations and trace calculus. \jmp{21}{1}, 170-174 (1980)

\noindent
\section*{1981-90}

\item
A. W. Weidemann. Quantum fields in a  ''Lorentz basis''. \nc {57A}{2}, 221-233
(1981)

\item
D. Han. The little group for photons and gauge transformations.
\ajp {49}{4}, 348-351  (1981)

\item
E. van der Spuy. The weak interaction related to the strong interaction. \nc
{71A}{3}, 305-332 (1982)

\item
R. Y. Levine and Y. Tomozawa. Supersymmetry and Lie algebras. \jmp {23}{8},
1415-1421 (1982)

\item
D. J. Almond. Off mass shell massless particles and the Weyl group in
 light cone coordinates. \jpa{15}{3}, 743-771  (1982)

\item
D. Han. Photon spin as a rotation in gauge space. \pr {D25}{2}, 461-463  (1982)

\item
V. V. Dvoeglazov and N. B. Skachkov. Three-dimensional covariant equation for a
composite system formed with a fermion and a boson (in Russian).
JINR Communications P2-84-199, Dubna,  Mar. 1984

\item
S. Weinberg. Causality, anti-particles and the spin statistics connection in
higher dimensions.
\pl {143B}{1-3}, 97-102 (1984)

\item
D. Han, Y. S. Kim and D. Son. Unitary transformations of photon polarization
vectors. \pr {D31}{2}, 328-330  (1985)

\item
N. Ohta. Causal fields and spin statistics connection for massless particles in
higher dimensions.
\pr {D31}{2}, 442-445  (1985)

\item
F. D. Santos. Relativistic spin-1 dynamics and deuteron-nucleus
elastic scattering.  \pl {175B}{2}, 110-114  (1986)

\item
L. Lukaszuk and L. Szimanowski. Tensor and bispinor representation of massless
fields.
\pr {D36}{8}, 2440-2457 (1987)

\item
J. S. Dowker. Vacuum averages for arbitrary spin around a cosmic string.
\pr {D36}{12}, 3742-3746 (1987)

\item
M. Daniel. On the fundamental spinor fields representing massless fermions
in $D$- dimensions. \pl {197B}{3}, 363-367 (1987)

\item
V. V. Dvoeglazov and N. B. Skachkov. Gluonium mass spectrum in the
quasipotential approach (in Russian). JINR Communications P2-87-882, Dubna,
Dec. 1987

\item
V. V. Dvoeglazov and N. B. Skachkov. Relativistic pa\-ra\-metriza\-tion of the
Hamiltonian
for the vector particle interaction with the external electromagnetic field.
\yf {48}{6}, 1770-1774 ; \sjnp {48}{6}, 1065-1068 (1988)

\item
M. Sivakumar. Consistent spin one theories by Kaluza-Klein dimensional
reduction.
\pr {D37}{6}, 1690-1693 (1988)

\item
F. D. Santos. Relativistic models of deuteron-nucleus scattering. In {\em Proc.
of the Workshop  on Relativistic Nuclear Many-Body Physics. Columbus, OH,
U.S.A., June 6-9, 1988.}\  Eds.
B. C. Clark, R. J. Perry and J. P. Vary. World Scientific, Singapore,  pp.
458-475 (1989)

\item
R. H. Good, jr. Relativistic wave equations for particles in
electromagnetic fields. \apny {196}{1}, 1-11 (1989)

\item
J. Jayaraman and M. A. B. de Oliveira. On the eigenvalues of $S \pi$ for
arbitrary spin
in a constant magnetic field. \jpa {22}{17}, 3531-3536  (1989)

\item
D. V. Ahluwalia and D. J. Ernst. Is light front quantum field theory merely a
change of coordinates? In {\em Proc.  of HUGS at CEBAF.  Hampton, VA, May
29-Jun 16, 1990,}
pp. 179-195

\noindent
\section*{1991-93}


\item
D. V. Ahluwalia and D. J. Ernst. Causal and relativistic propagator for spin-2
hadrons. Preprint
CTP-TAMU-28-91, Texas  A\& M, 1991

\item
D. V. Ahluwalia and D. J. Ernst. Relativistic phenomenology of high spin
hadrons. 1. Covariant
spinors and  causal propagators. Preprint CTP-TAMU-57-91, Texas A\& M, 1991

\item
V. V. Dvoeglazov. Quantum field description of interactions between two
relativistic
particles having various spins (in Russian). Cand. Sci. (Ph. D.) dissertation
JINR 2-91-331,  Dubna,  Jul. 1991

\item
D. V. Ahluwalia. Relativistic quantum field theory of high spin matter fields:
a pragmatic approach for hadronic physics. Ph. D. thesis UMI-92-06454-mc, Texas
A\& M University,
Aug. 1991

\item
F. D. Santos and H. van Dam. Relativistic dynamics of spin-one particles and
deuteron-nucleus scattering. \pr {C34}{1}, 250-261 (1991)

\item
D. V. Ahluwalia and D. J. Ernst. On kinematical acausality in Weinberg's
equations
for arbitrary spin. In {\em Classical and Quantum Systems: Foundations and
Symmetries.
Proc. of the 2nd Int. Wigner Symp. Groslar, Germany, July 16-20, 1991.}

\item
D. V. Ahluwalia. Instant form $\rightarrow$ front form: collapse of spinorial
degrees of
freedom from four $\rightarrow$ two and related  matters. Preprint
CTP-TAMU-87-91, Texas
A \& M, Oct. 1991

\item
V. K. Mishra {\it et al.} Implications of various spin-one relativistic wave
equations for intermediate-energy deuteron-nucleus scattering. \pr {C43}{2},
801-811 (1991)

\item
A. Amorim  and F. D. Santos. From Dirac phenomenology to deuteron-nucleus
elastic scattering at intermediate energies. \pr {C44}{5}, 2100-2110 (1991)

\item
D. V. Ahluwalia and D. J. Ernst. Phenomenological approach to high-spin quantum
fields based on Weinberg formalism. \pr {C45}{6}, 3010-3012 (1992)

\item
D. V. Ahluwalia and D. J. Ernst. Conceptual framework for high spin hadronic
physics.
In {\em AIP Conf. -- Proceedings of the Computational Quantum Physics
Conference. Nashville,
TN, May 23-25, 1991.} Vol. 260,  pp. 309-315 (1992)

\item
D. V. Ahluwalia and D. J. Ernst.  Paradoxical kinematic acausality in
Weinberg's equations for massless particles of arbitrary spin.
\mpl {A7}{22}, 1967-1974 (1992)

\item
D. V. Ahluwalia. Interpolating Dirac spinors between instant and light front
forms.
\pl {B277}{3}, 243-248 (1992)

\item
D. V. Ahluwalia and D. J. Ernst.  New arbitrary-spin wave equations
for $(j,0)\oplus (0,j)$ matter fields without kinematic acausality
and constraints. \pl {B287}{1}, 18-22 (1992)

\item
L. P. Horwitz and N. Shnerb. On the group theory of the polarization states of
a gauge field.  Preprint IASSNS-92-56, Princeton, Sept. 1992

\item
V. Bhansali. How the little group constrains massless field representations in
higher
even dimensions.  Preprint HUTP-92-A052 (hep-th/9209128), Harvard, Oct. 1992

\item
D. V. Ahluwalia and M. Sawicki.  Natural hadronic  degrees of freedom for an
effective QCD action in the  front form. Preprint LA-UR-92-3133
(hep-ph/9210204), Los Alamos, 1992

\item
D. V. Ahluwalia, D. J. Ernst and  C. Burgard. Dirac-like relativistic
phenomenology
of high spin hadrons. 2. Kinematic structure of the Weinberg equations.
Preprint
93-0167, Los Alamos, 1993

\item
V. V. Dvoeglazov and S. V. Khudyakov. Vector particle interactions in
the quasipotential approach. Preprint IFUNAM FT-93-019 (hep-ph/9306246),
Mexico,
Jun. 1993

\item
D. V. Ahluwalia, T. Goldman and M. B. Johnson. Majorana-like $(j,0)\oplus
(0,j)$
representation spaces: construction and physical interpretation. Preprint
LA-UR-93-2645
(hep-th/9307118), Los Alamos, Jul. 1993

\item
M. Pillin. $Q$- deformed relativistic wave equations. Preprint MPI-PH-93-61
(hep-th/9310097), M\"unich, Jul. 1993

\item
V. V. Dvoeglazov, S. V. Khudyakov and S. B. Solganik.  Relativistic covariant
equal time
equation for quark-diquark system. Preprint  IFUNAM FT-93-024 (hep-th/9308305),
Mexico, Aug. 1993

\item
D. V. Ahluwalia and T. Goldman. Space-time symmetries and vortices in the
cosmos. \mpl {A8}{28}, 2623-2630  (1993) (hep-ph/9304242)

\item
D. V. Ahluwalia and M. Sawicki. Front form spinors in the Weinberg-Soper
formalism
and generalized Melosh transformations for any spin. \pr {D47}{11}, 5161-5168
(1993)

\item
D. V. Ahluwalia and D. J. Ernst.  $(j,0)\oplus (0,j)$ covariant spinors and
causal propagators based on Weinberg formalism. \ijmp {E2}{2}, 397-422 (1993)

\item
D. V. Ahluwalia, T. Goldman and M. B. Johnson. A Bargmann-Wightman-Wigner type
quantum field theory. \pl {B316}{1}, 102-108  (1993) (hep-ph/9304243)

\item
V. V. Dvoeglazov. Lagrangian formulation of the  $2(2S+1)$ -- component  model
and its connection with the skew-symmetric tensor description. \hj{16}{6},
459-467 (1993) (hep-th/9305141)

\item
V. V. Dvoeglazov. Interactions of fermions with massless
spin-one particles in the $2(2S+1)$- component
formalism. \hj{16}{5}, 423-428 (1993) (hep-th/9306108)

\item
V. V. Dvoeglazov and S. V. Khudyakov. Gluonium as bound state of massive gluons
described by the Joos-Weinberg wave functions. Preprint IFUNAM FT-93-35
(hep-ph/9311347), Mexico, Nov. 1993

\item
D.V. Ahluwalia, M.B. Johnson, T. Goldman. Space-time symmetries: $P$ and $CP$
violation.
Preprint LA-UR-93-4314 (hep-th/9312089), Los Alamos, Sep. 1993

\item
D.V. Ahluwalia, T. Goldman, M.B. Johnson. $(J,0)\oplus (0,J)$ representation
space: Dirac-like construct. Preprint LA-UR-93-4315 (hep-th/9312090), Los
Alamos, Dec. 1993

\item
D.V. Ahluwalia, T. Goldman, M.B. Johnson. $(J,0)\oplus (0,J)$ representation
space: Majorana-like construct. Preprint LA-UR-93-4316 (hep-th/9312091), Los
Alamos, Dec. 1993

\item
M. Sawicki and D. V. Ahluwalia. Parity transformation in the front form.
Preprint LA-UR-93-4317
(hep-th/9312092), Los Alamos, Dec. 1993

\end{enumerate}
}

\begin{thebibliography}{99}
\normalsize{
\bibitem{PDG} Particle Data Group, {\it Review  of Particle properties.}
{\it Phys. Rev. D}{\bf 45 (part~II)} (1992) \\[-7mm]
\bibitem{rar} W. Rarita and J. Schwinger, {\it Phys. Rev.} {\bf 60} (1941)
61\\[-7mm]
\bibitem{barg} V. Bargmann and E. P. Wigner, {Proc. Natl. Acad. Sci. (USA)}
{\bf 34} (1948) 211\\[-7mm]
\bibitem{Joos} H. Joos, {\it Forts. Phys.} {\bf 10} (1962) 65\\[-7mm]
\bibitem{Weinberg} S. Weinberg, {\it Phys. Rev.} {\bf 133} (1964) B1318; ibid
{\bf 134} (1964) B882; ibid {\bf 181} (1969) 1893.\\[-7mm]
\bibitem{ahl} D. V. Ahluwalia and D. J. Ernst, {\it Phys. Lett. B}{\bf 287}
(1992) 18; {\it Phys. Rev. C}{\bf 45} (1992) 3010; {\it Mod. Phys. Lett. A}{\bf
7} (1992) 1967\\[-7mm]
\bibitem{ahl2} D. V. Ahluwalia and T. Goldman, {\it Mod. Phys. Lett. A}{\bf 8}
(1993)  2623; D. V. Ahluwalia, T. Goldman and  M. B. Johnson, {\it  Phys. Lett.
B}{\bf 316} (1993) 102;  Preprints  LA-UR-93-2645, 4314, 4315, 4316, Los
Alamos, 1993\\[-7mm]
\bibitem{Bargm} E. P. Wigner, in {\it Group Theoretical Concepts and Methods in
Elementary Particle Physics -- Lectures of the Istanbul Summer School of
Theoretical Physics, 1962.} Ed. F. G\"ursey\\[-7mm]
\bibitem{Hatf} B. F. Hatfield {\it Quantum Field Theory of Point and  Strings.}
Addison-Wesley, Rendwood, ~U.~S.~A.,~1992\\[-7mm]
\bibitem{Corson} E. M. Corson, {\it Introduction to Tensors, Spinors and
Relativistic
Wave Equations. Relation Structure.} New York, Chelsea, 1982\\[-7mm]
\bibitem{tirring} W. Thirring and H. Narnhofer,  Preprint UWThPh-1991-63
(mp-arc/93-11), Wien, 1991; H. Huffel, {\it Phys. Lett. B}{\bf 296} (1992) 233
\\[-7mm]
\bibitem{Hos} J. Hosek, Preprint JINR  E2-82-542, Jul. 1982; Preprint
CERN-TH-4104/85, Feb 1985; {\it Phys. Rev. D}{\bf 36} (1987) 2093 \\[-7mm]
\bibitem{sohrtoki}  D. G. Boulware, {\it Ann. Phys.} {\bf 56} (1970) 140; A.
Salam and J. Strathdee, {\it Phys. Rev. D}{\bf 2}  (1970) 2869.\\[-7mm]
\bibitem{Arno} M. Sheinblatt and R. Arnowitt, {\it Phys. Rev. D}{\bf 1} (1970)
1603\\[-7mm]
\bibitem{tam} A. C. Tam and W. Happer, {\it Phys. Rev. Lett.} {\bf 38} (1977)
278\\[-7mm]
\bibitem{Naik} P. C. Naik and T. Pradhan, {\it J. Phys. A}{\bf 14} (1981) 2795;
T. Pradhan, R. P. Malik and P. C. Naik, {\it Pramana J. Phys.} {\bf 24} (1985)
77\\[-7mm]
\bibitem{Nemes} M. C. Nemes {\it et al.}, Preprints HEP-TH 9307068, 9307069,
9307070, 9309155, 9311106.\\[-7mm]
\bibitem{pr11} G. S. Adkins, A. A. Salahuddin and K. E. Schalm, {\it Phys. Rev.
A}{\bf 45} (1992) 3333; ibid, 7774\\[-7mm]
\bibitem{pg21} C. I. Westbrook, D. W.Gidley, R. S. Conti and A. Rich,  {\it
Phys. Rev. A} {\bf 40} (1989) 5489;  J. S. Nico, D. W. Gidley, A. Rich and P.
W. Zitzewitz, {\it Phys. Rev. Lett.} {\bf 65} (1990) 1344\\[-7mm]
\bibitem{sam} M. A. Samuel and G. Li, Preprint SLAC-PUB-6318, Stanford,
1993\\[-7mm]
\bibitem{Weaver} D. L. Weaver, C. L. Hammer  and R. H. Good, jr., {\it Phys.
Rev.}  {\bf
135}  (1964) B241\\[-7mm]
\bibitem{barut} A. O. Barut, I. Muzinich and D. N. Williams, {\it Phys. Rev.}
{\bf 130} (1963) 442\\[-7mm]
\bibitem{mat} P. M. Mathews, {\it Phys. Rev.} {\bf 143} (1966) 978\\[-7mm]
\bibitem{will} S. A. Williams, J. P. Draayer  and  T. A. Weber, {\it Phys.Rev.}
 {\bf 152} (1966) 1207\\[-7mm]
\bibitem{Sankaranarayanan} A. Sankaranarayanan and R. H. Good, Jr., {\it Nuovo
Cimento} {\bf 36} (1965) 1303; A. Sankaranarayanan, {\it Nuovo Cimento} {\bf
38} (1965) 889\\[-7mm]
 \bibitem{Good}  D. Shay and R. H. Good , jr.,  {\it Phys. Rev.}  {\bf  179}
(1969) 1410; L. D. Krase, Pao  Lu and R. H. Good, jr.,  {\it Phys. Rev. D}{\bf
3} (1971) 1275\\[-7mm]
\bibitem{Tucker} R. H. Tucker and C. L. Hammer, {\it Phys. Rev. D}{\bf 3}
(1971) 2448\\[-7mm]
\bibitem{Shay} D. Shay, H. S. Song and R. H. Good, jr., {\it Nuovo Cimento
Suppl.} {\bf 3} (1965) 455\\[-7mm]
\bibitem{pursey} C. L. Hammer, S. C. McDonald and D. L. Pursey, {\it Phys.
Rev.} {\bf 171} (1968) 1349\\[-7mm]
\bibitem{Recami2} E. Majorana, {\it Scientific manuscripts} (1928-32),
as reported E. Recami, R. Mignani and M. Baldo,  {\it Lett.
Nuovo Cim.} {\bf 11} (1974) 568\\[-7mm]
\bibitem{Chow} T. L. Chow, {\it J. Phys. A}{\bf 14} (1981) 2173\\[-7mm]
\bibitem{Gianetto} E. Gianetto, {\it  Lett. Nuovo Cim.} {\bf 44} (1985)
140, 145\\[-7mm]
\bibitem{Ryder} L. H. Ryder, {\it Quantum Field Theory.} Cambridge Univ. Press,
1985\\[-7mm]
\bibitem{Dvoeglaz} V. V.  Dvoeglazov,  {\it Hadronic J.} {\bf 16} (1993) 423;
459; Preprints IFUNAM  FT-93-19, 24, 35,  Mexico, 1993; V. V. Dvoeglazov  and
N. B. Skachkov, JINR Communications  P2-84-199, P2-87-882,  Dubna, 1987;  {\it
Sov. J. Nucl. Phys.} {\bf 48} (1988) 1065\\[-7mm]
\bibitem{Hayashi} K. Hayashi, {\it Phys. Lett. B}{\bf 44} (1973) 497.\\[-7mm]
\bibitem{Lipkin} D. M.  Lipkin , {\it J. Math. Phys.} {\bf 5} (1964) 696; A.
Sudbery,
{\it J. Phys. A}{\bf 19} (1986) L33.\\[-7mm]
\bibitem{Marinov} M. S.  Marinov,  {\it Ann. Phys.} {\bf 49}  (1968)
357\\[-7mm]
\bibitem{Moshinsky} M. Moshinsky M. and A. Szcepaniak,  {\it J. of Phys. A}{\bf
22}  (1989)  L817\\[-7mm]
\bibitem{Dixit} V. V. Dixit, T. S. Santhanam and W. D. Thacker, {\it J. Math.
Phys.} {\bf 33} (1992) 1114\\[-7mm]
\bibitem{berest} V. B. Berestetskii, E. M. Lifshitz and L. P. Pitaevskii, {\it
Relativistic Quantum Theory. Part I.} Pergamon Press, 1979\\[-7mm]
\bibitem{Jancewicz} B. Jancewicz,  {\it Multivectors and
Clifford Algebra in Electrodynamics.} World Sci. Singapore, 1988\\[-7mm]
\bibitem{a11}   N. B. Skachkov,  JINR Communications  P2-12152, Dubna, 1979;
JINR Preprints  E2-81-294, E2-81-308, E2-81-399, Dubna, 1981;
N. B. Skachkov and I. L. Solovtsov, {\it Sov. J. Part. and Nucl.} {\bf 9}
(1978) 1\\ [-7mm]
\bibitem{Shap} I. S. Shapiro, {\it Sov. Phys. Doklady} {1} (1956) 91; {\it Sov.
Phys. JETP} {\bf 43} (1962) 1727; {\it Phys. Lett.} {\bf 1} (1962) 253}
\end{thebibliography}
\end{document}